\documentclass[twocolumn]{aastex631}

\usepackage{color}
\usepackage{fancybox}
\usepackage{newtxtext,newtxmath}

\begin{document}

\title{Universal multi-stream radial structures of cold dark matter halos}

\thisfancyput(14.5cm,0.5cm){\large{YITP-22-163}}

\author{Yohsuke Enomoto}
\affiliation{Department of Physics, Kyoto University \\
Kyoto 606-8502, Japan}

\author[0000-0002-9664-0760]{Takahiro Nishimichi}
\affiliation{Center for Gravitational Physics and Quantum Information, Yukawa
Institute for Theoretical Physics, Kyoto University, Kyoto 606-8502, Japan}
\affiliation{Kavli Institute for the Physics and Mathematics of the Universe (WPI),
The University of Tokyo Institutes for Advanced Study, The University of Tokyo, 5-1-5 Kashiwanoha, Kashiwa, Chiba 277-8583, Japan}
\affiliation{Department of Astrophysics and Atmospheric Sciences, Faculty of Science,
Kyoto Sangyo University, Motoyama, Kamigamo, Kita-ku, Kyoto 603-8555, Japan}

\author[0000-0002-4016-1955]{Atsushi Taruya}
\affiliation{Center for Gravitational Physics and Quantum Information, Yukawa
Institute for Theoretical Physics, Kyoto University, Kyoto 606-8502, Japan}
\affiliation{Kavli Institute for the Physics and Mathematics of the Universe (WPI),
The University of Tokyo Institutes for Advanced Study, The University of Tokyo, 5-1-5 Kashiwanoha, Kashiwa, Chiba 277-8583, Japan}



\begin{abstract}
Virialized halos of cold dark matter generically exhibit multi-stream structures of accreted dark matter within an outermost radial caustic known as the splashback radius. 
By tracking the particle trajectories that accrete onto the halos in cosmological $N$-body simulations, we count their number of apocenter passages ($p$), and use them to characterize the multi-stream structure of dark matter particles. 
We find that the radial density profile for each stream, classified by the number of apocenter passages, exhibits universal features, and can be described by a double power-law function comprising inner shallow and outer steep slopes of indices of $-1$ and $-8$, respectively. Surprisingly, these properties hold over a wide range of halo masses. 
The double-power law feature is persistent when dividing the sample by concentration or accretion rate. 
The dependence of the characteristic scale and amplitude of the profile on $p$ cannot be replicated by known self-similar solutions, requiring consideration of complexities such as the distribution of angular momentum or mergers.
\end{abstract}
%%%%%%%%%%%%%%%%%%%%%%%%%%%%%%%%%%%%%%%%%%%%%%%%%%%%%%
%%%%%%%%%%%%%%%%%%%%%%%%%%%%%%%%%%%%%%%%%%%%%%%%%%%%%%



%%%%%%%%%%%%%%%%%%%%%%%%%%%%%%%%%%%%%%%%%%%%%%%%%%%%%%
%%%%%%%%%%%%%%%%%%%%%%%%%%%%%%%%%%%%%%%%%%%%%%%%%%%%%%
\section{Introduction} \label{sec:intro}
%%%%%%%%%%%%%%%%%%%%%%%%%%%%%%%%%%%%%%%%%%%%%%%%%%%%%%
%%%%%%%%%%%%%%%%%%%%%%%%%%%%%%%%%%%%%%%%%%%%%%%%%%%%%%

Since its first indication by Zwicky \citep{Zwicky1933,Zwickly1937}, dark matter has long been thought to be an essential ingredient to explain the cosmic structure formation across a wide range of observations. One important consequence, supported by various observations, is that dark matter forms,  at the macroscopic level, a smooth distribution with virtually null initial local velocity dispersion, referred to as cold dark matter (CDM) \citep{Peebles1982,Peebles1984,Blumenthal_etal1984}. 
According to a widely accepted scenario, a collapse of CDM occurs within
a cosmic web, leading to the formation of self-gravitating bound objects called dark matter halos. The late-time evolution of such halos is driven by the continuous accretion of surrounding dark matter onto the halo center and successive mergers with other halos, resulting in the structure of CDM halos exhibiting several characteristic features. 
One prominent feature, found in cosmological $N$-body simulations, is the so-called Navarro-Frenk-White (NFW)   profile, first suggested by \citet{Navarro_Frenk_White1996,Navarro_Frenk_White1997}, which has had a significant impact on observations as a testing ground for the CDM paradigm. Another striking feature is the power-law nature of the pseudo-phase-space density profile defined by $Q(r)\equiv\rho(r)/\sigma^3(r)$, with $\rho(r)$ and $\sigma(r)$  being  respectively the radial density and velocity  dispersion profile  \citep{Taylor_Navarro2001,Navarro_etal2010,Ludlow_etal2010}. The slope found in simulations closely matches the  prediction of the Bertschinger’s secondary infall model \citep{Bertschinger1985}, 
indicating that the structure of halos is built up in a self-similar manner. These properties imply that there is something more fundamental underlying them, which could aid in understanding the origin of universalities. 

In this {\it Letter}, we present an alternative way to characterize halo density profiles based on phase-space information. It is worth noting that the formation of CDM halos is accompanied by
a shell crossing at an early phase of matter accretion, 
followed by a multi-stream motion of matter distribution. The multi-stream nature of halos has garnered recent attention, highlighted with a renewed interest, 
as the outer boundary of the multi-stream region 
serves as a natural boundary of halos, and  
is clearly demarcated by a radial caustic, manifested as a local steepening of the density profile, referred to as the splashback radius \cite[e.g.,][]{Diemer_Kravtsov2014,Adhikari_Dalal_Chamberlain2014,More_Diemer_Kravtsov2015}.
Motivated by these findings, \cite{Sugiura20} developed a method using an extension of the SPARTA algorithm in \cite{Diemer2017} to reveal the multi-stream nature of halos at the outer regions and they found that about $30$\% of halos are well-described by the self-similar solution of \citet{Fillmore_Gorldreich1984}. In this {\it Letter}, by substantially refining their analysis based on high-resolution simulations with finely sampled snapshots out to an early 
halo formation, we are able to unveil the innermost parts of the multi-stream region, where we find that halos exhibit a universal feature in each multi-stream distribution.   

%%%%%%%%%%%%%%%%%%%%%%%%%%%%%%%%%%%%%%%%%%%%%%%%%%%%%%
%%%%%%%%%%%%%%%%%%%%%%%%%%%%%%%%%%%%%%%%%%%%%%%%%%%%%%
\section{Method} \label{sec:method}
%%%%%%%%%%%%%%%%%%%%%%%%%%%%%%%%%%%%%%%%%%%%%%%%%%%%%%
%%%%%%%%%%%%%%%%%%%%%%%%%%%%%%%%%%%%%%%%%%%%%%%%%%%%%%

We analyze cosmological $N$-body simulations performed in a flat $\Lambda$CDM cosmology, which is consistent with recent observations of cosmic microwave background radiation~\citep{Planck2015cosmology}. We mainly analyze the simulation that follows the movements of $500^3$ particles in a comoving box with a side length of $41\,h^{-1}\mathrm{Mpc}$ using the Tree Particle-Mesh code \textsc{Ginkaku}~(Nishimichi, Tanaka \& Yoshikawa, in preperation). We employ a softening length of $4.1\,h^{-1}\mathrm{kpc}$, which we denote by $r_\mathrm{LR}$ in what follows. The snapshots of the particles are saved at $1,001$ redshifts, evenly spaced between $z=0$ and $5$, providing a dense sampling to accurately determine the number of apocenter passages (denoted by $p$ in what follows) up to $\sim 50$, following the method of \citet{Sugiura20} with minor modifications. 

We first select relaxed halos from those identified by \textsc{Rockstar}~\citep{Rockstar} at $z=0$, by imposing a cut in the spin parameter and the offset between the center of mass and the density peak~\citep{Klypin16}. We also discard subhalos according to the consistency between the exact spherical-overdensity mass and that listed in the Rockstar catalog. We then trace the main progenitor by following the particles within the virial radius back in time, updating the center and the list of member particles using the shrinking-sphere method at each snapshot until we reach the first snapshot at $z=5$ or the number of member particles falls below $1,000$. Our final halo trajectories are defined as the center of mass of the $1,000$ fixed member particles, which are closest in phase space to the center of the main progenitor at the highest redshift to which we can trace the progenitor with at least $1,000$ particles. We next follow forward in time the center of mass of these fixed particles to obtain a smooth trajectory robust to merger events. We 
monitor the velocities and positions of all surrounding particles that are within $2.5 \, R_\mathrm{vir}$ at $z=0$ relative to the center of the progenitor. We 
define and count the apocenter passage
for each particle when the relative velocity changes from outgoing to infalling and the relative position has orbited at least $90^\circ$ from the previous apocenter passage~\citep{Sugiura20}. These specific choices are found to be robust for the determination of the number of apocenter passages up to $\sim 50$.

%%%%%%%%%%%%%%%%%%%%%%%%%%%%%%%%%%%%%%%%%%%%%%%%%%%%%%
\begin{figure}[t!]
\epsscale{1.2}
\plotone{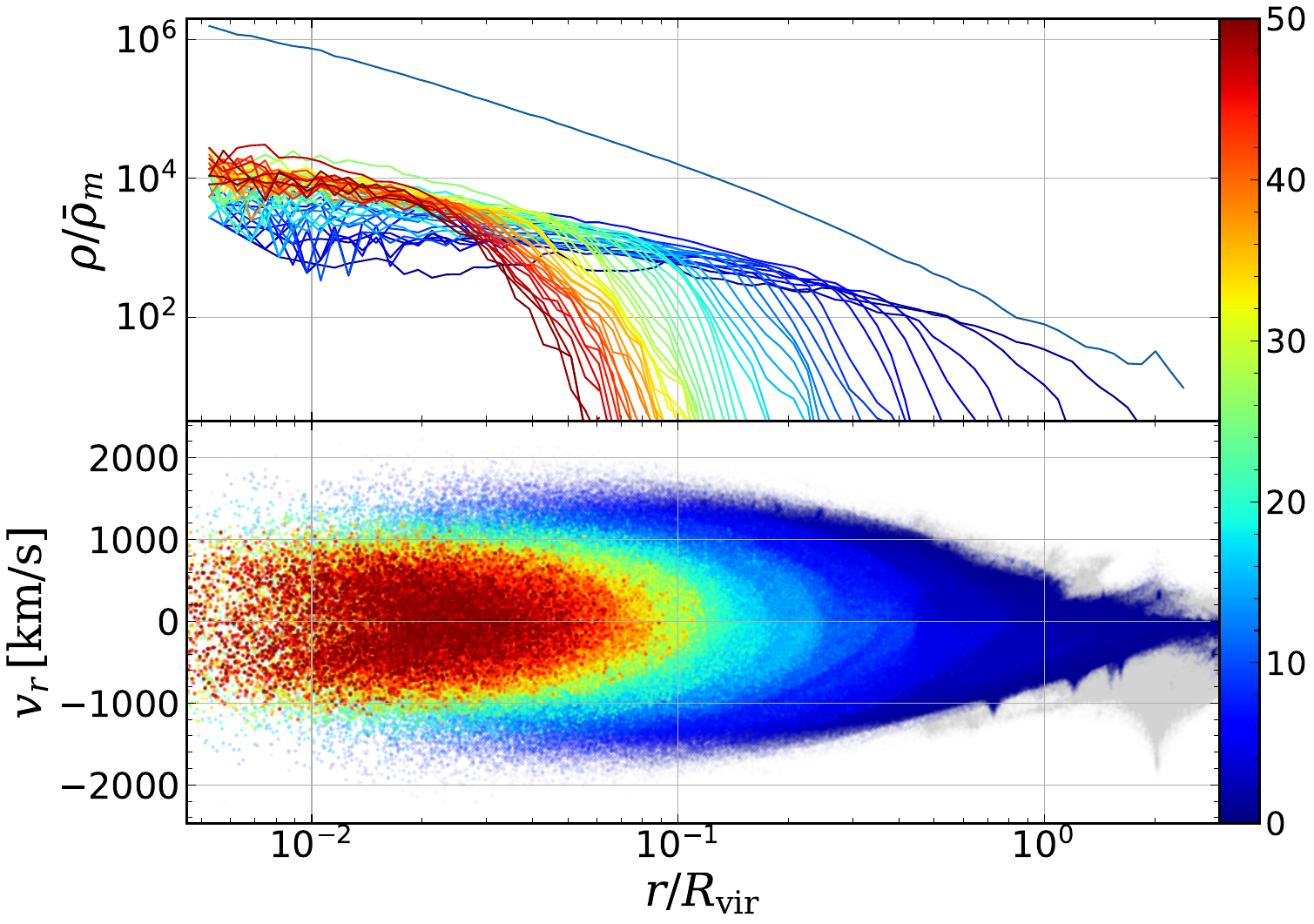}
\caption{Radial density profile (upper) and phase-space distribution (lower) of a halo with $M_\mathrm{vir}=1.49\times 10^{14}\,h^{-1}M_\odot$. The upper panel shows the decomposition of the total density profile (highest line) into the contributions from $N$-body particles with different numbers of apocenter passages, represented by colors ranging from $p=1$ (dark blue) to $p=50$ (dark red). 
The results are all normalized by the background matter density of the universe, $\overline{\rho}_{\rm m}$.
The lower panel displays the distribution of individual particles, with the same color coding. The infalling component, $p=0$, is depicted in gray.  
\label{fig:indivhalo}}
\end{figure}

%%%%%%%%%%%%%%%%%%%%%%%%%%%%%%%%%%%%%%%%%%%%%%%%%%%%%%

In Fig.~\ref{fig:indivhalo}, we present the radial density profile and phase-space distribution of a representative halo with mass $M_{\rm vir}=1.49\times10^{14}\,h^{-1}\,M_\odot$, color-coded by the number of apocenter passages, $p$. It is apparent that particles with a high value of $p$ tend to be concentrated at smaller radii, leading to an increase in density 
and a reduction in radial velocity dispersion, resulting in an onion-like multi-stream structure in the phase-space distribution. 
We also confirmed that tangential velocity dispersion decreases with increasing $p$.
On the other hand, the density profiles exhibit similar features, with the inner and outer slopes converging to a specific value regardless of $p$. 
In the following sections, we will further analyze this behavior for halos with different properties.

In order to study the convergence, we have conducted a higher-resolution simulation with $2,000^3$ particles with an identical initial Gaussian random field. However, storing as many as $\gtrsim 1,000$ snapshots from this simulation requires a significant amount of disk space, and 
an accurate apocenter count would be costly. Therefore, we only use this run to verify the density profile at $z=0$. 
In the following discussion, we refer to this simulation as HR, while the one with $500^3$ particles is called LR. 
The softening scale for HR is $r_\mathrm{HR} = 1.025\,h^{-1}\mathrm{kpc}$.

%%%%%%%%%%%%%%%%%%%%%%%%%%%%%%%%%%%%%%%%%%%%%%%%%%%%%%
%%%%%%%%%%%%%%%%%%%%%%%%%%%%%%%%%%%%%%%%%%%%%%%%%%%%%%
\section{Results} \label{sec:results}
%%%%%%%%%%%%%%%%%%%%%%%%%%%%%%%%%%%%%%%%%%%%%%%%%%%%%%
%%%%%%%%%%%%%%%%%%%%%%%%%%%%%%%%%%%%%%%%%%%%%%%%%%%%%%

\begin{figure*}[ht!]
\includegraphics[width=\linewidth]{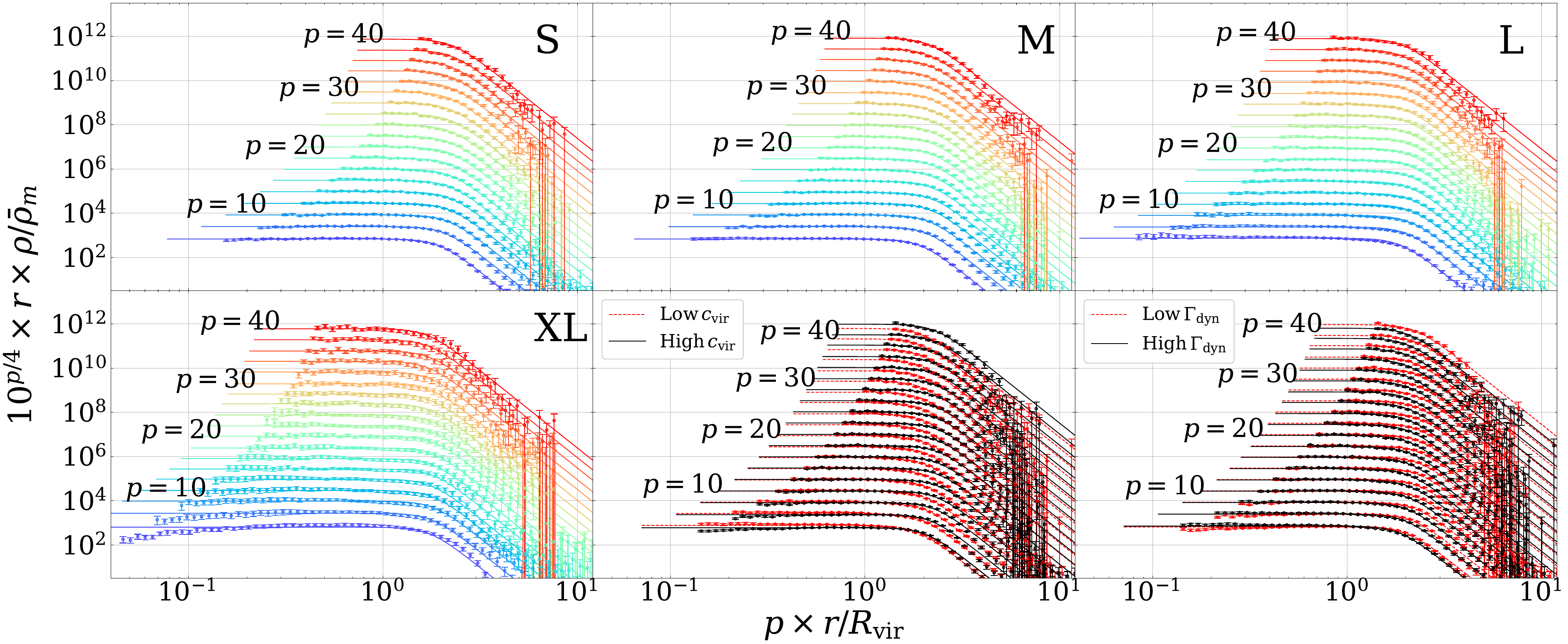}
\caption{Stacked radial density profiles of $N$-body particles with even number of apocenter passages, ranging from $p=4$ to $40$. 
The four mass bins are displayed in the upper three and the lower left panel for S, M, L and XL, respectively.
Additionally, the lower middle and right panels show the results obtained from $460$ halos in the mass range $[4.10\time10^{11},\,2.30\times10^{12}]\,h^{-1}M_\odot$, %dividing them 
which are further divided
into the two subsamples based on the concentration parameter $c_{\rm vir}$ and accretion rate $\Gamma_{\rm dyn}$, respectively (see text in detail). In each panel, the fitted results with Equation~(\ref{eq:double_powerlaw_fit}) are depicted as solid lines.}
\label{fig:general}
\end{figure*}

\begin{figure}[ht!]
\epsscale{1.15}
\plotone{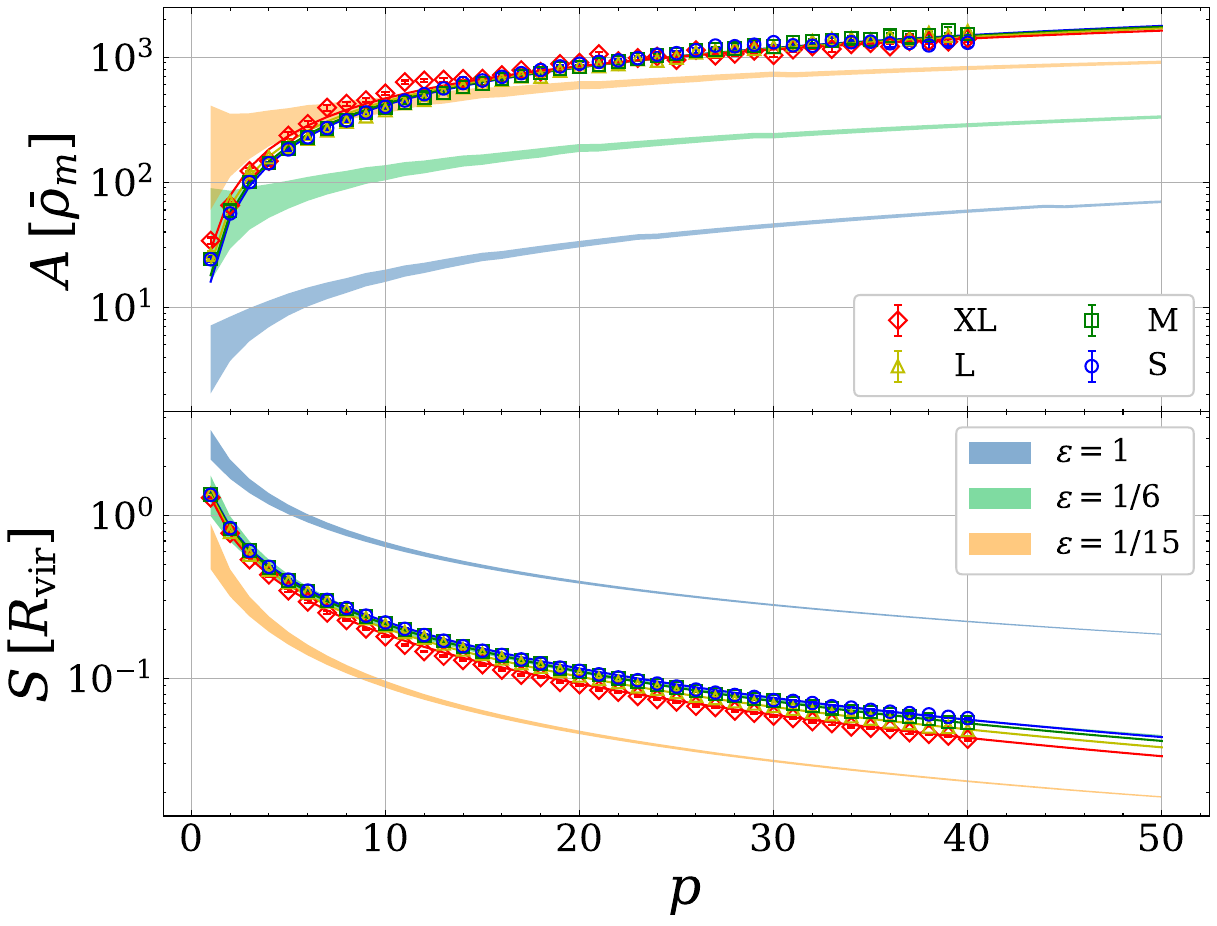}
\caption{
Dependence of the characteristic density $A$ (upper) and scale $S$ (lower) on the number of apocenter passages, $p$, as determined by fitting to Equation~(\ref{eq:double_powerlaw_fit}) in different symbols four mass bins (see legend). The thin solid curves represent the fitting formulae, Eqs.~(\ref{eq:fitA}) and (\ref{eq:fitS}). For comparison, predictions of the Fillmore-Goldreich self-similar solutions are also shown, 
for specific values of the parameter
$\epsilon$ ($1/15$, $1/6$ and $1$). 
In plotting these predictions, we identify the position of radial caustics in the self-similar solutions with the characteristic scale $S(p)$, and derive $A(p)$ by equating the masses contained in each stream. The shaded regions for the predictions indicate uncertainty in identifying $S(p)$ with the position of the $p$-th or $(p+1)$-th radial caustics of the self-similar solutions.}
\label{fig:fitparams}
\end{figure}

In order to systematically and quantitatively investigate the radial density profile of particles with $p$ apocenter passages, we divide the halos into four mass bins (Table \ref{tab:halo_mass_bin}). For each mass bin, we measure the stacked profile for each stream by rescaling the radial coordinate by the virial radius and the density by the virial mass for individual halos. 
We 
find that the stacked profile for each $p$, as presented in the upper panels and the lower left panel in Figure~\ref{fig:general}, are well-fitted by the following functional form\footnote{
The fitting analysis employs the standard $\chi^2$ method, with weights determined by the inverse variance among the stacked halos. Radial bins with $r > 1.2\, r_\mathrm{LR}$ are considered.}
:
%%%%%%%%%%%%%%%%%%%%%%%%%%%%%%%%%%%%%%%%%%%%%%%%%%%%%%
\begin{equation}
        \rho_{\rm stream}(r;p)=\frac{A(p)}{\{r/S(p)\} 
        \bigl[1+\{r/S(p)\}^7\bigr]},
    \label{eq:double_powerlaw_fit}
\end{equation}
%%%%%%%%%%%%%%%%%%%%%%%%%%%%%%%%%%%%%%%%%%%%%%%%%%%%%%
where the characteristic scale $S(p)$ and density $A(p)$ are given as a function of $p$. Due to the rescaling in the stacking, the functions, $S(p)$ and $A(p)$, are quite similar among the four mass bins, as shown in Figure~\ref{fig:fitparams}. 
We find that these functions are well captured by
%%%%%%%%%%%%%%%%%%%%%%%%%%%%%%%%%%%%%%%%%%%%%%%%%%%%%%
\begin{equation}\label{eq:fitA}
\begin{split}
    \log_{10}{\Bigl\{A_\mathrm{fit}(p)/\overline{\rho}_{\rm m}\Bigr\}}  = 4.89-0.119\log_{10}{\left(M_\mathrm{vir,10}\right)}  \\ +\Bigl\{-3.89+0.243\log_{10}{\left(M_\mathrm{vir,10}\right)}\Bigr\}\,p^{-9/40},
\end{split}
\end{equation}
%%%%%%%%%%%%%%%%%%%%%%%%%%%%%%%%%%%%%%%%%%%%%%%%%%%%%%
%%%%%%%%%%%%%%%%%%%%%%%%%%%%%%%%%%%%%%%%%%%%%%%%%%%%%%
\begin{align}\label{eq:fitS}
\begin{split}
    \log_{10}{\Bigl\{S_\mathrm{fit}(p)/R_{\rm vir}\Bigr\}} = 
    2.46-0.0474\log_{10}{\left(M_\mathrm{vir,10}\right)}  \\ +
    \Bigl\{-2.29-0.0639\log_{10}{\left(M_\mathrm{vir,10}\right)}\Bigr\}\,p^{1/8},
\end{split}
\end{align}
%%%%%%%%%%%%%%%%%%%%%%%%%%%%%%%%%%%%%%%%%%%%%%%%%%%%%%
including the weak mass dependence,
where $M_\mathrm{vir,10}$ is defined by $M_\mathrm{vir}/10^{10}h^{-1}M_\odot$.

In Figure~\ref{fig:general}, the stacked profiles are presented up to $p=40$.
The horizontal axis is scaled by $(p/R_\mathrm{vir})$ and the vertical axis by $(10^{p/4}\,r/\overline{\rho}_{\rm m})$ for improved visibility.
The errorbars indicate the standard error of the stacked profiles.
The profiles are in agreement with Equation~(\ref{eq:double_powerlaw_fit}) over a wide range of radii and mass scales. A sharp cutoff is observed in the profile at large radii, which is consistent with our model with the asymptotic slope of $-8$ (see also \citet{2022arXiv220503420D} for an alternative characterization of the outer cutoff for orbiting particles). 
More notably, the inner slope tends to be consistent with $-1$ for most cases, except for orbits with $p\lesssim10$ for the XL sample, which exhibit a shallower slope.
This is likely due to the sensitivity of these low-$p$ orbits to recent mass accretion or merger history~\citep[e.g., ][]{Sugiura20}. However, this trend tends to be erased after several orbits, reaching a universal slope for $p\gtrsim10$, indicating a self-similar growth of phase-space structure.

\begin{figure*}[ht!]
\epsscale{1.15}
\plotone{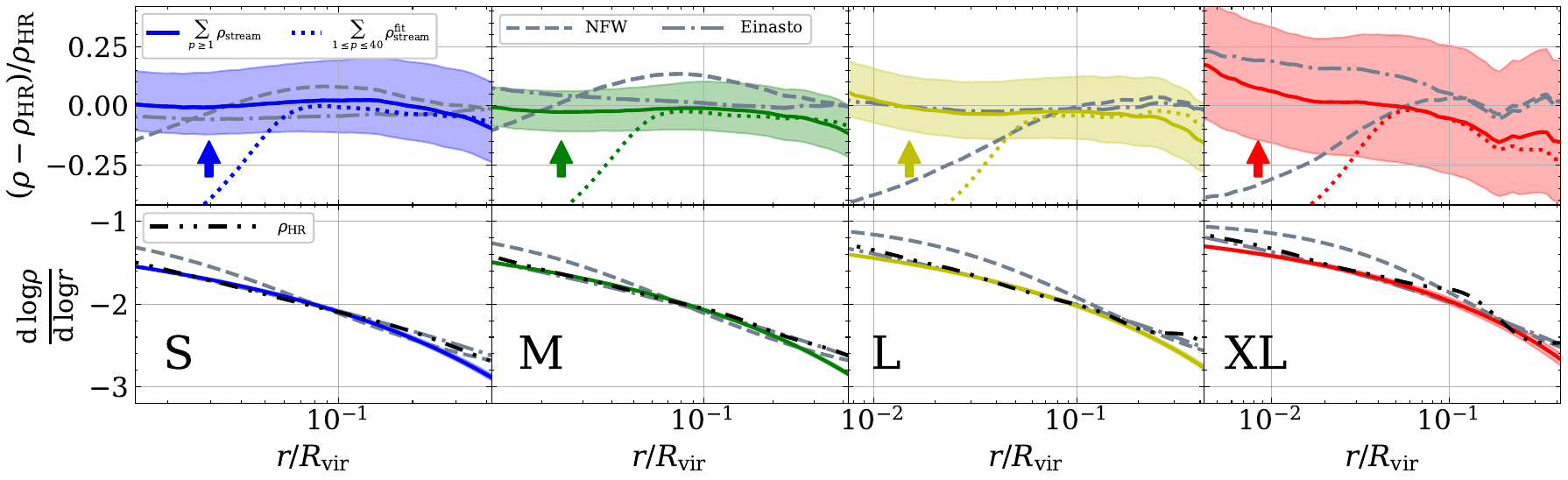}
\caption{
Comparison of the total density profile between our model
($\sum\rho_{\rm stream}$) and HR ($\rho_\mathrm{HR}$). The upper and lower panels, respectively, show the fractional difference with respect to HR, i.e., $(\rho-\rho_{\rm HR})/\rho_{\rm HR}$, and the logarithmic slope, $d\log\rho/d\log r$. 
The results for four mass bins are presented separately in each panel for scales above $2\,{\rm Max}(r_{\rm HR}/R_{\rm vir})$, i.e., twice the maximum value of the ratio $r_{\rm HR}/R_{\rm vir}$ estimated for individual halos in each mass bin.
The shaded regions indicate the estimated uncertainties in the prediction, which are propagated from the statistical error in the stacked profile through the uncertainties in the fitting parameters.
We also plot the NFW (dashed) and Einasto (dotted-dashed; \protect\citet{einasto1965}) profiles, obtained by fitting HR in the range $2 \, {\rm Max}(r_{\rm HR}/R_{\rm vir}) \leq r/R_{\rm vir}\leq 0.9$. 
In the upper panels, the results obtained from a partial summation of the double power-law profile up to $p=40$ are also plotted (dotted). 
The vertical arrows indicate the scale of $1.2\,{\rm Max}(r_{\rm LR}/R_{\rm vir})$, corresponding to the convergence radius above which measured profiles from LR and HR simulations agree well with each other at $\sim3\%$ precision.}
\label{fig:fitsum}
\end{figure*}

To quantitatively assess the double power-law nature of each stream, we 
compare the total density profile from HR for 
halos that have been matched with LR to the prediction obtained by summing the individual double power-law profiles described by Equations~(\ref{eq:double_powerlaw_fit})--(\ref{eq:fitS})\footnote{In the plot, the summation is conservatively taken up to $p=3,000$. The change in density is less than $0.2\%$ over the plotted range when we instead stop at $p=300$.}. 
The results are shown in Figure \ref{fig:fitsum}, where the solid lines with shaded regions represent the prediction based on the double power-law model, taking into account the uncertainties in the numerical coefficients in the fit.
Our model is in good agreement with HR for all four mass bins. Notably, we can recover the profile even at 
$r/R_{\rm vir}\leq 1.2\,{\rm Max}(r_\mathrm{LR}/R_{\rm vir})$
despite the fact that the individual profiles for each $p$ are fitted to the scales larger than $1.2\,{\rm Max}(r_\mathrm{LR}/R_{\rm vir})$ and only up to $p=40$.
This suggests that the model effectively extrapolates the mass distribution to large values of $p$ beyond the resolution limit of LR. In the lower panel, we can observe the transition of the slope from $-3$ to $-1$ in different models\footnote{The logarithmic slope is estimated from discrete simulation data points with statistical noise using \texttt{GEORGE} python package~\citep{GEORGE} for Gaussian Process.}.

%%%%%%%%%%%%%%%%%%%%%%%%%%%%%%%%%%%%%%%%%%%%%%%%%%%%%%
%%%%%%%%%%%%%%%%%%%%%%%%%%%%%%%%%%%%%%%%%%%%%%%%%%%%%%

%%%%%%%%%%%%%%%%%%%%%%%%%%%%%%%%%%%%%%%%%%%%%%%%%%%%%%
%%%%%%%%%%%%%%%%%%%%%%%%%%%%%%%%%%%%%%%%%%%%%%%%%%%%%%

%%%%%%%%%%%%%%%%%%%%%%%%%%%%%%%%%%%%%%%%%%%%%%%%%%%%%%
%%%%%%%%%%%%%%%%%%%%%%%%%%%%%%%%%%%%%%%%%%%%%%%%%%%%%%

%%%%%%%%%%%%%%%%%%%%%%%%%%%%%%%%%%%%%%%%%%%%%%%%%%%%%%
\begin{figure*}[ht!]
\epsscale{1.15}
\plotone{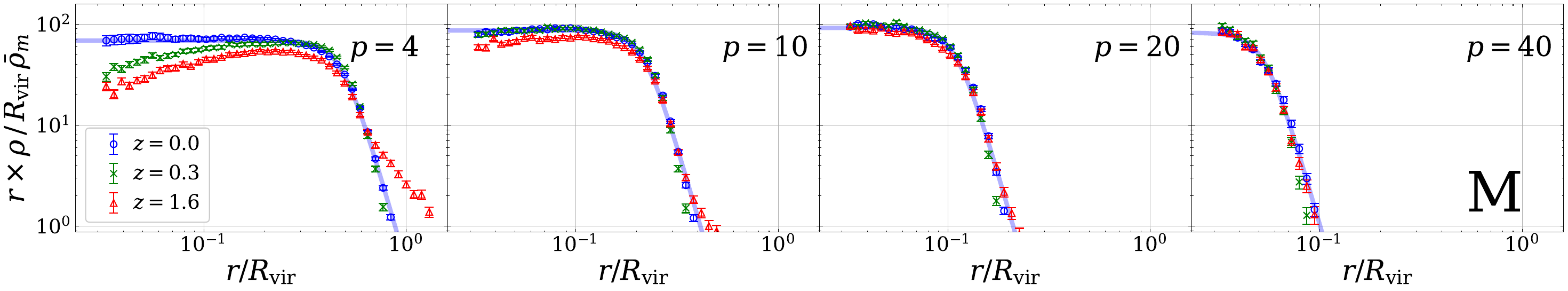}
\caption{Stacked density profiles for the halo sample M measured at $z=0$ (blue), $0.3$ (green) and $1.6$ (red). From left to right panels, results for the particles with $p=4$, $10$, $20$, and $40$ are respectively shown.
Note that in each panel, the number of apocenter passages $p$ is the one measured at $z=0$, and we simply trace back the particles identified at $z=0$ to higher redshifts $z=0.3$ and $1.6$. 
Each profile is computed in physical units (not comoving) and normalized by the virial radius $R_\mathrm{vir}$ (horizontal axis) and background mass density $\bar{\rho}_{\rm m}$ (vertical axis) at $z=0$. 
For ease of comparison, we also multiply the resultant density by $r$. 
The errorbars indicate the standard error of the stacked profiles.
}
\label{fig:tevolution}
\end{figure*}
%%%%%%%%%%%%%%%%%%%%%%%%%%%%%%%%%%%%%%%%%%%%%%%%%%%%%%
%
%%%%%%%%%%%%%%%%%%%%%%%%%%%%%%%%%%%%%%%%%%%%%%%%%%%%%%
\begin{table}[t!]
  \caption{Halo samples in the LR simulation. The second, third, and fourth columns respectively show the range of halo masses, virial radii, and the number of halos.}
  \label{tab:halo_mass_bin}
  \centering
  \begin{tabular}{cccc}
  \hline\hline
    Sample & %$[M_{\rm min},\;M_{\rm max}]\;h^{-1}M_\odot$ 
    $10^{11}\,M_\mathrm{vir}\,[h^{-1}M_\odot]$
    & 
    % $[R_{\rm min},\;R_{\rm max}]\;h^{-1}\mathrm{kpc}$ 
    $R_\mathrm{vir}\,[h^{-1}\mathrm{Mpc}]$
    & \# of halos \\ 
    \hline
    % S &  $[3.16,\;5.71]$ & $[139,\;169]$& 300 
    S &  $[3.16,\;5.71]$ & $[0.14,\;0.17]$& 300 
    \\
%    \hline
    % M & $[5.71,\;24.2]$ &$[169,\;272]$& 300 
    M & $[5.71,\;24.2]$ &$[0.17,\;0.27]$& 300 
    \\
%    \hline
    % L & $[24.2,\;134]$ &$[273,\;475]$& 70  
    L & $[24.2,\;134]$ &$[0.27,\;0.48]$& 70  
    \\
%    \hline
    % XL & $[134,\;1530]$ &$[483,\;1076]$& 13 
    XL & $[134,\;1530]$ &$[0.48,\;1.08]$& 13 
    \\
    \hline
      \end{tabular}
\end{table}
%%%%%%%%%%%%%%%%%%%%%%%%%%%%%%%%%%%%%%%%%%%%%%%%%%%%%%
%%%%%%%%%%%%%%%%%%%%%%%%%%%%%%%%%%%%%%%%%%%%%%%%%%%%%%

%%%%%%%%%%%%%%%%%%%%%%%%%%%%%%%%%%%%%%%%%%%%%%%%%%%%%%
%%%%%%%%%%%%%%%%%%%%%%%%%%%%%%%%%%%%%%%%%%%%%%%%%%%%%%
\section{Discussion} \label{sec:discussion}
%%%%%%%%%%%%%%%%%%%%%%%%%%%%%%%%%%%%%%%%%%%%%%%%%%%%%%
%%%%%%%%%%%%%%%%%%%%%%%%%%%%%%%%%%%%%%%%%%%%%%%%%%%%%%

%----------------------------------------------------%
%----------------------------------------------------%
%\subsection{Universality in biased subsamples}
\subsection{Dependence on halo samples}
\label{subsec:halo_samples}
%----------------------------------------------------%
%----------------------------------------------------%

The remarkable double power-law features in section \ref{sec:results} are seen in mass-selected halo samples. Here, to assess the robustness of our findings, we analyze a subset of $460$ halos within a specific mass range $[4.10\times10^{11},\,2.39\times10^{12}]\,h^{-1}M_\odot$.
These halos are divided into two sub-samples based on two different criteria.
We employ the concentration parameter $c_{\rm vir}$, defined by the ratio $R_{\rm vir}/R_{\rm s}$ with $R_{\rm s}$ being the scale radius of the NFW profile, and the mass accretion rate defined by $\Gamma_{\rm dyn}(t)\equiv \{\log[M(t)]-\log[M(t-t_{\rm dyn})]\}/\{ \log[a(t)]-\log[a(t-t_{\rm dyn})]\}$ with $t_{\rm dyn}$ being the dynamical time estimated from halo masses \citep{Diemer2017}\footnote{
We use the virial mass, $M_\mathrm{vir}$, to measure $\Gamma_\mathrm{dyn}$, whereas \cite{Diemer2017} uses $M_\mathrm{200m}$.}. 
Note that the radius $R_{\rm s}$ is estimated in
\textsc{Rockstar} based on the maximum circular velocity \citep{Klypin_etal2011}. In both cases,
we divide the halos into two halves, one with high values of these indicators and the other with low values.

The middle bottom (right bottom) panel of Figure \ref{fig:general} depicts the results for two subsamples having low and high values of $c_{\rm vir}$ ($\Gamma_{\rm dyn}$), represented by red and black colors, respectively. Again, a good agreement between the double power-law function and measured profiles is observed over a wide range of $p$. A close look at each stream profile reveals that halos with high concentration or low accretion rate tend to have a large amplitude $A(p)$ and a large characteristic scale $S(p)$. These trends are particularly evident for larger $p$, suggesting that the universal double power-law feature is established in a self-regulated manner during the orbital motion in the multi-stream region, where the diversity of mass accretion and merger histories tend to be erased and only be imprinted in $A(p)$ and $S(p)$.

%----------------------------------------------------%
%----------------------------------------------------%
\subsection{Comparison with self-similar solutions}
\label{subsec:self-similar_solutions}
%----------------------------------------------------%
%----------------------------------------------------%

The results in section \ref{sec:results} and \ref{subsec:halo_samples} strongly indicate that the inner structure of halos is built up dynamically in a self-similar manner. 
Here, we compare our results with self-similar solutions.
While self-similar solutions are only valid in the Einstein-de Sitter universe, the secondary infall model of \cite{Bertschinger1985} has been shown to reproduce the pseudo phase-space density of $Q(r)\propto r^{-1.875}$ found in simulations in the $\Lambda$CDM model. 
Along the line of this, we consider the spherically symmetric solutions put forth by \citet{Fillmore_Gorldreich1984}, which include the Bertschinger's secondary infall model as a special case.
Recent work by \citet{Sugiura20} has made a direct comparison of these predictions with radial multi-stream structures obtained from simulations up to $p=5$. Identifying the position of radial caustics in self-similar solutions with the characteristic scale of the double power-law profile in Equation~(\ref{eq:double_powerlaw_fit}), it is possible to make predictions for both $A(p)$ and $S(p)$. 

In Figure \ref{fig:fitparams}, we compare the predictions of self-similar solutions with our $N$-body results for three values of the model parameter $\epsilon$, which describes the power-law slope of initial density contrast.
Note that the parameter $\epsilon$ is restricted to the range $[0,\,1]$, and the solution with $\epsilon=1$ corresponds to Bertschinger's secondary infall model. Figure \ref{fig:fitparams} shows that none of the solutions consistently explain the trends in both $A(p)$ and $S(p)$, although setting the parameter $\epsilon$ to $1/6$ reproduces the characteristic scale $S(p)$ reasonably well. 
The main reason for this failure is that for each stream, the Fillmore-Goldreich solutions predict a 
steep inner profile with a logarithmic slope of around $-2$ irrespective of the value of $\epsilon$. 
One possible explanation for the shallow inner cusps found in simulations is to introduce the non-zero angular momentum, which can reduce the steepness of the profile near the halo center~\citep{2001MNRAS.325.1397N,2010PhRvD..82j4044Z}. 
However, existing solutions allow for the introduction of angular momentum in a very specific manner, and without a broad angular momentum distribution, they fail to describe the shallow inner cusp seen in the profile for each $p$. 

We thus conclude that a more comprehensive theoretical study is needed to fully understand the universal features found in this \textit{Letter}, taking into account the complexities associated with mass accretion and merger history.
This may involve exploring the angular momentum distribution or relaxing the symmetry assumptions ~\citep[see][for the latter aspect]{1993ApJ...418....4R,Lithwick_2011}.

%----------------------------------------------------%
%----------------------------------------------------%
\subsection{On the emergence of double-power law nature}
\label{subsec:}
%----------------------------------------------------%
%----------------------------------------------------%

As a final discussion toward a better understanding of the origin of the universal double power-law nature, we focus on the halo sample M in Table \ref{tab:halo_mass_bin}, and select the particles with $p=4$, $10$, $20$, and $40$ at $z=0$. Then, we trace back their trajectories to higher redshifts and measure the density profiles for each value of $p$ stacked over different halos. Figure~\ref{fig:tevolution} overplots the results at $z=0.3$ (green) and $1.6$ (red), on top of those at $z=0$ already shown in Figure \ref{fig:general} (black). Clearly, the profiles vary over redshifts, and the amplitude of density gets increased as decreasing $z$. 
Interestingly, however, the evolution of the inner profiles becomes significantly weaker as the value of $p$ increases, and at $p=40$, the profiles almost converge even at the outer most part. 
This suggests that the double power-law nature was established at an early stage of the halo formation and remains stable against matter accretion, which can only affect the outer part of the density profile represented by particles with small values of $p$.
Apart from the origin of the universal profiles, this picture is consistent with previous studies that show that the accreting matter mainly piles up at the outer region \citep[e.g., ][]{Zhao.2003}, and partly explains why the characteristic scale $S(p)$ in Equation (\ref{eq:double_powerlaw_fit}) is a decreasing function of $p$; particles with large $p$ have accreted earlier and their distribution tends to be relaxed at the inner part of halos. 
In this respect, the dynamics at the early stage of halo formation would be the key to clarifying the origin of the double power-law nature.

%%%%%%%%%%%%%%%%%%%%%%%%%%%%%%%%%%%%%%%%%%%%%%%%%%%%%%
\section{Conclusion} \label{sec:conclusion}
%%%%%%%%%%%%%%%%%%%%%%%%%%%%%%%%%%%%%%%%%%%%%%%%%%%%%%
%%%%%%%%%%%%%%%%%%%%%%%%%%%%%%%%%%%%%%%%%%%%%%%%%%%%%%
In this \textit{Letter}, we have investigated the multi-stream radial structures of dark matter halos in cosmological $N$-body simulations. Our focus is on the radial distribution of dark matter particles within the splashback radius. We use the method developed by \cite{Sugiura20} to trace the trajectories of dark matter particles and quantify the density profile for each stream, which we label by $p$. With the help of $1,001$ snapshots between $z=0$ and $5$, we are able to resolve the multi-stream structure in phase space up to $p=40$.
The radial density profiles for each stream are accurately described by a double power-law function (Equation~\ref{eq:double_powerlaw_fit}), with characteristic density $A(p)$ and scale $S(p)$ well-fitted respectively to Equations ~\eqref{eq:fitA} and ~\eqref{eq:fitS}. 
These results are consistent across different sample selections based on the concentration parameter and mass accretion rate. 
We can recover the total density profile by summing up the individual contribution modeled by Equation ~(\ref{eq:double_powerlaw_fit}), which provides a prediction comparable to or even better than the Einasto profile. 
Our findings suggest that the double power-law nature seen in the stream profiles is universal. 
This remarkable characteristic appears to have been established during an early stage of matter accretion and remains stable.
To gain a deeper understanding of these results, we compare them with predictions based on self-similar solutions. We find that the Fillmore-Goldreich solutions (nor Bertschinger's solution as a special case) cannot consistently explain both $A(p)$ and $S(p)$. This suggests that a more comprehensive theoretical study is necessary, taking into account the dynamical complexities associated with halo accretion and merging history.

The universal features of halos found in this \textit{Letter} are a direct consequence of the cold nature of dark matter and serve as valuable insights into the physical properties of CDM halos. 
While this study has utilized $N$-body simulations and investigated the inner multi-stream structure up to $p=40$, 
recent developments in simulating collisionless self-gravitating systems through
Vlasov-Poisson equations offer a promising way to further probe the phase-space structure ~\citep{Yoshikawa_Yoshida_Umemura2013, Hahn_Angulo2016, Sousbie_Colombi2016}. 
This would provide a deeper understanding of the physics behind the universal features.
To search for observational evidence of this universality, it would also be beneficial to investigate the impact of baryonic feedback through hydrodynamical simulations.

Finally, another point worth further investigating is to scrutinize the radial phase-space structures for alternative dark matter models, as the nature of dark matter 
has a significant impact on small-scale structure formation~\citep[e.g.,][for a review]{Bullock_Boylan-Kolchin2017}. 
Our method to reveal multi-stream structures can be straightforwardly applied to simulations of other dark matter models. 
Any difference in the radial multi-stream structures could provide valuable observational
probes to clarify the nature of dark matter.

%%%%%%%%%%%%%%%%%%%%%%%%%%%%%%%%%%%%%%%%%%%%%%%%%%%%%%
\begin{acknowledgments}
We thank St\'ephane Colombi and Takashi Hiramatsu for insightful suggestions and discussions, Shogo Ishikawa and Satoshi Tanaka for comments and discussions. This work was
supported in part by MEXT/JSPS KAKENHI Grant
Number JP19H00677 (TN), JP20H05861, JP21H01081
(AT and TN), and JP22K03634 (TN). We also acknowledge financial support from Japan Science and Technology Agency (JST) AIP Acceleration Research Grant Number JP20317829
(AT and TN). 
YE is also supported by JST, the establishment of university fellowships towards the creation of science technology innovation, Grant Number JPMJFS2123.
Numerical computations were carried out 
at Yukawa Institute Computer Facility, and Cray XC50 at
Center for Computational Astrophysics, National Astronomical Observatory of Japan.
\end{acknowledgments}
%%%%%%%%%%%%%%%%%%%%%%%%%%%%%%%%%%%%%%%%%%%%%%%%%%%%%%

% \vspace{5mm}

% \software{Anything else to put here ?}
%           }

% \appendix

% \section{Appendix information}

%\begin{thebibliography}{99}
%\end{thebibliography}
\bibliography{references}{}
\bibliographystyle{aasjournal}

\end{document}